# On the effectiveness of local vortex identification criteria in the compressed representation of wall-bounded turbulence

Chengyue Wang[1,a)], Qi Gao[2], Biao Wang[1,b)]

**AFFILIATIONS**

[1] Sino-French Institute of Nuclear Engineering and Technology, Sun Yat-sen University, Zhuhai, China

[2] State Key Laboratory of Fluid Power and Mechatronic System, Department of Mechanics, School of Aeronautics and Astronautics, Zhejiang University, Hangzhou, China

[a)]**Electronic mail**: wang.chengyue@163.com
[b)]**Electronic mail**: wangbiao@mail.sysu.edu.cn

**ABSTRACT**

Compressing complex flows into a tangle of vortex filaments is the basic implication of the classical notion of the vortex representation. Various vortex identification criteria have been proposed to extract the vortex filaments from available velocity fields, which is an essential procedure in the practice of the vortex representation. This work focuses on the effectiveness of those identification criteria in the compressed representation of wall-bounded turbulence. Five local identification criteria regarding the vortex strength and three criteria for the vortex axis are considered. To facilitate the comparisons, this work first non-dimensionalize the criteria of the vortex strength based on their dimensions and root mean squares, with corresponding equivalent thresholds prescribed. The optimal definition for the vortex vector is discussed by trialling all the possible combinations of the identification criteria for the vortex strength and the vortex axis. The effectiveness of those criteria in the compressed representation is evaluated based on two principles: (1) efficient compression, which implies the less information required, the better for the representation; (2) accurate decompression, which stresses that the original velocity fields could be reconstructed based on the vortex representation in high accuracy. In practice, the alignment of the identified vortex axis and vortex isosurface, and the accuracy for decompressed velocity fields based on those criteria are quantitatively compared. The alignment degree is described by using the differential geometry method, and the decompressing process is implemented via the two-dimensional field-based linear stochastic estimation. The results of this work provide some reference for the applications of vortex identification criteria in wall-bounded turbulence.

## I. INTRODUCTION

Extracting vortex structures from available velocity fields is a routine procedure when analyzing the physics of complex flows. Although a widely-accepted mathematical definition for vortex is not available yet, the concept typically refers to the rotating motion of a multitude of material particles around a common centre.[1] This notion of vortex brings two aspects of convenience for the corresponding analysis work. First, as an intensely rotational motion, a strong vortex retains coherence for comparatively larger spatial separations and longer temporal intervals, which implies a significant contribution to the transfer of mass and momentum. Second, the vortex cores or the vortex skeletons typically presents a tube-like shape and thus can be simplified as a curve with variable radius and strength, which benefits the vortex-based modelling works. As Perry and Chong[2] stated, "these vortex skeletons are the 'genetic code' of the flow field, since this requires very little specification and the Biot-Savart law can be used to generate the velocity field."

Plenty of research works focus on the vortex identification criteria, and these works have been comprehensively reviewed by Chakraborty et al.[3], Kolář[4] and Epps[5]. Chakraborty et al.[3] classified the identification criteria into two types: the local approaches based on velocity gradient tensor ($\nabla \mathbf{u}$) and the non-local ones. In this work, only the local identification criteria are discussed, which includes the discriminant $\Delta$,[6] the second invariant $Q$,[7] the complex part of the eigenvalues $\lambda_{ci}$ of $\nabla \mathbf{u}$,[8] and the second largest eigenvalue $\lambda_2$ for pressure Hessian ($\nabla\nabla p$).[9] These criteria are mostly employed as scalar indicators for the local vortex strength, and the vortex cores can be represented by the isosurfaces of these criteria for certain prescribed thresholds. Kolář[4] argued that the vortex identification criterion should recognize the vortex axis as well as the vortex strength and he developed a triple decomposition technique regarding $\nabla \mathbf{u}$ in order to meet this requirement. In fact, as reminded by Cucitore et al.[10], some of the foregoing scalar criteria also imply privileged directions in their definitions, which are correlated with the vortex axis. The $\lambda_2$ criterion recognizes the eigendirection $\Lambda_p$ associate with the minimum eigenvalue of pressure Hessian as the vortex axis.[9] The $\lambda_{ci}$ criterion involves two privileged directions: the real eigenvector $\Lambda_r$ of $\nabla \mathbf{u}$ which indicates the compressing or stretching direction of the local flow, and the direction perpendicular to the real part or the imaginary part of the complex eigenvector $\Lambda_{cr} + i\Lambda_{ci}$, which refers to the swirl-plane-normal direction (i.e. $\Lambda_{cr} \times \Lambda_{ci}$).[8] These indicating vectors for the vortex axis have been employed in the investigation of wall-bounded turbulence. Zhou[11] found the good alignment of $\Lambda_r$ with the vortex axis by tracking the $\Lambda_r$ vectors in a recognized hairpin vortex core. Pirozzoli et al.[12] compared $\Lambda_r$, $\Lambda_{cr} \times \Lambda_{ci}$, and $\Lambda_p$, and found they are visually equivalent for vortices placed sufficiently far away from the wall. Gao et al.[13] employed $\Lambda_r$ to identify the vortex axis and refined the vortex strength, density and orientation based on planar particle image velocimetry (PIV) data and DNS data. Wang et al.[14] extended the work of Gao et al.[13] for higher Reynolds numbers based on the volumetric PIV data and analyzed the role of $\Lambda_r$ in controlling the evolution of the vorticity.

Recently, Liu et al.[15] and Tian et al.[16] proposed a new identification criterion involving both the vortex strength and the vortex axis, which triggered widespread interest in the physical interpretation and application of the new method.[17-20] Basically, the new method considers the rotation state of the local material line segment and recognizes a direction without rotation as the vortex axis. The vortex strength is defined as twice the minimum angular speed in the plane normal to the vortex axis, which was named "Rortex" by Liu et al.[15] Coincidentally, the definition of Rortex is consistent with several prevenient research works. While the magnitude of Rortex is equivalent to the residual vorticity in the triple decomposition of Kolář[4] for two-dimensional (2D) situations,[21] the orientation of Rortex was demonstrated to be $\Lambda_r$.[20] Tian et al.[22] revisited the definition of Rortex by analyzing the physical meaning of the local rotation of fluid elements.

Although these identification criteria share the same purpose of extracting vortex cores, quantitatively comparing and ranking their effectiveness is impossible unless the specific applying situation is confined. While Jeong and Hussain[9] tried to show the superiority of the $\lambda_2$ criterion, Cucitore et al.[10] argued that any of the well-known criteria has counter-examples where unsatisfying identification results are obtained. Chakraborty et al.[3] demonstrated that these identification criteria (not including the newly proposed Rortex) are compatible with one another for typical turbulent flows by using the equivalent thresholds. Chen et al.[23] compared various criteria for planar velocity fields in wall-bounded turbulence, and investigated equivalent thresholds in order to facilitate quantitative comparison of the results from different criteria. Gao and Liu[20] showed the advantage of Rortex in the robustness to the addictive shear flows compared to other eigenvalue-based criteria. Zhan et al.[24] compared the $Q$ criterion and Rortex in the application of an in-stream structure and concluded Rortex is more suitable in their analysis. Kolář and Šístek[21] investigated the stretching response of Rortex and other criteria and found that both Rortex and $\lambda_{ci}$ are stretching-insensitive schemes, and they allow an arbitrary axial strain.

This work will focus on the application of the vortex identification criteria in wall-bounded turbulence. The effectiveness of the identification criteria for the vortex strength ($\Delta$, $Q$, $\lambda_2$, $\lambda_{ci}$, and $R$) and the vortex axis ($\Lambda_{cr} \times \Lambda_{ci}$, $\Lambda_r$ and $\Lambda_p$) will be quantitatively compared from the perspective of the compressed representation (CR) of wall-bounded turbulence. CR is a basic idea for turbulence modelling, and it means representing the turbulent flows as simplified vortex skeletons. The famous attached eddy theory could be viewed as the successful practice of CR, which has achieved promising predictions for the wall-bounded turbulence.[25] Typically, two procedures are necessary for the practice of CR: identifying vortex cores or skeletons from the velocity fields, and accurately reconstructing the velocity fields based on the modelled vortices, which is much like the "compression" and "decompression" of turbulent velocity fields.[3]

The vortex identification criterion plays a vital role in the first procedure of CR for wall-bounded turbulence. The technique for extracting the vortex centerlines based on the identified results has been developed by Zhu and Xi[26], which provides good representations of original flows. Besides, the direct results for identification criteria (such as the $\lambda_{ci}$ field) could also be viewed as CRs for wall-bounded turbulence, since the high-intensity regions for the criteria are very sparse and the isolated structures usually take simple tube-like shapes. From the perspective of turbulence modelling, the isolated structures should bring the necessary information for inferring the original velocity fields, which could be viewed as the "gene code" of the turbulent flows as the quotation from Perry and Chong[2]. Based on these considerations, the performance of identification criteria could be evaluated from two aspects: the efficiency in turbulence data compression and the accuracy of the corresponding decompression. While the former aspect emphasizes that the less information required, the better for the identified results, the later aspect stresses that the compression should be invertible in the sense that the original velocity fields can be reconstructed accurately based on the compressed representation.

The effectiveness of vortex identification criteria in the efficient compression aspect could be evaluated by checking the alignment of the identified vortex isosurface and vortex axis. An ideal criterion should return tube-like vortex structures, where the vortex axis should be tangent to the local isosurface and align with the axial direction of the vortex tube. The alignment benefits the simplifying and modelling works for vortex skeletons, which will be elaborated in Sec. III. In this work, this alignment will be quantitatively estimated by using the differential geometry method, and all the combinations of the criteria for the vortex strength and the vortex axis will be investigated in the application of wall-bounded turbulence.

The effectiveness of vortex identification criteria in the accurate decompression aspect could be accessed by the correlation coefficients for the ground-truth DNS velocity fields and reconstructed ones based on identified vortices. In this work, the vortex-to-velocity (V2V) reconstruction could be implemented by a newly-proposed data-driven method, namely, field-based linear stochastic estimation (FLSE).[27] For numerical convenience, this work only employs the 2D version of FLSE, which could also be viewed as the expansion of the one-dimensional (1D) spectral linear stochastic estimation.[28] The implementation scheme and results of FLSE in this work will be introduced in detail.

The following part of this work will be arranged as follows. In Sec. II, the DNS data employed in this work and a collection of the definitions for vortex identification criteria will be provided. Particularly, the threshold equivalence will be focused in order to facilitate the following comparing works. Subsequently, in Sec. III, the effectiveness of the criteria in the efficient compression aspect will be evaluated by investigating the alignment of the identified vortex isosurface and vortex axis. In Sec. IV, the effectiveness regarding the accurate decompression aspect will be evaluated. 2D FLSE will be employed to reconstruct the velocity fields based on the vortex fields defined by various criteria, and the corresponding reconstruction accuracy will be compared. Lastly, the conclusions will be provided in Sec. V.

## II. DNS DATA AND THE VORTEX IDENTIFICATION CRITERIA
### A. DNS data

The data employed in this investigation come from an open-access direct numerical simulation (DNS) database for turbulent boundary layers (Fluid Dynamics Group of Universidad Politecnica de Madrid, available at "https://torroja.dmt.upm.es/"). Details for the DNS code and the validation works can be found in the papers of Borrell et al.[29] and Sillero et al.[30] The DNS data have been used in several research works, including Marusic et al.[31], Wang et al.[32], Wang et al.[33], and Wang et al.[14]. The whole DNS data correspond to a developing turbulent boundary layer, containing 15361, 4096 and 535 collocation points for the streamwise, spanwise and wall-normal direction, respectively. In this work, only the data segment for $Re_\tau \approx 1200$ is truncated from the whole DNS data. The truncated DNS segment has a dimension of $8\delta \times 13.4\delta \times 0.3\delta$ ($\delta$ is the averaged boundary thickness), which corresponds to $1412 \times 4096 \times 96$ calculation nodes. The wall-normal range considered is $0 \sim 0.3\delta$, covering the whole buffer layer, the whole logarithmic layer, and part of the wake layer. The separation between adjacent collocation points along the wall-normal direction is not uniform, which is determined by the resolution of the spatial discretization scheme and the local Kolmogorov scale.[29] The streamwise spacing and the spanwise spacing are 6.80 and 3.93 wall-units. In the following discussion, let the three coordinating axes of $x, y, z$ be aligned with the streamwise, spanwise and wall-normal direction. $u, v, w$ denote the three fluctuating velocity components, respectively. A superscript of '+' indicates the quantity normalized based on the local wall-unit or the friction velocity. Following Pirozzoli et al.[12], all the vortex identification criteria are calculated based on the fluctuating velocity components, and thus only $u, v, w$ are considered in this work. The three-component fluctuating velocity field and its gradient tensor are denoted as $\mathbf{u}$ and $\nabla\mathbf{u}$, respectively.

### B. Identification criteria for vortex strength and equivalent thresholds

Definitions of the identification criteria regarding the vortex strength involved in this work are collected in Table I. As we can see, while $\lambda_2$ considers eigenvalues of pressure Hessian, all the other criteria are rooted in the physical interpretation of $\nabla\mathbf{u}$. According to Chakraborty et al.[3], $\Delta$ and $Q$ can be determined by the complex eigenvalue ($\lambda_{cr} + i\lambda_{ci}$) of $\nabla\mathbf{u}$, which is listed in the fourth column of this table. A similar formula for $\lambda_2$ is not available unless $\nabla\mathbf{u}$ is a normal tensor. The explicit expression for $R$ is provided by Wang et al.[34], which reveals a direct relation to $\lambda_{ci}$ and the vorticity component along $\mathbf{\Lambda}_r$ (i.e. $\omega_r$). All these formulas for the identification criteria contain $\lambda_{ci}$, which seems to indicate that $\lambda_{ci}$ is closely correlated to all the other criteria.

The effective ranges for these criteria are shown in the third column of the table, which implicate the most basic requirements when employed to identify vortices. Referring to formulas in the fourth columns, we can see these effective ranges are not equivalent to one another. While $Q > 0$ yields $\lambda_{ci} > 0$ and $|\lambda_{cr}/\lambda_{ci}| < \sqrt{3}/3$, $\lambda_2 < 0$ demands $\lambda_{ci} > 0$ and $|\lambda_{cr}/\lambda_{ci}| < 1$ (only when $\nabla\mathbf{u}$ is a normal tensor). The ratio $\lambda_{cr}/\lambda_{ci}$ is the inverse spiraling compactness, which measures the local orbital compactness in a vortex.[3] The effective ranges for the other three criteria (i.e. $\Delta > 0$, $\lambda_{ci} > 0$ and $R > 0$) are mathematically equivalent. They do not involve the reqirement of spiraling compactness and impose comparatively looser requirements in terms of the effective ranges.

These criteria have different dimensions: $\Delta$ has the dimension of $\omega^6$ ($\omega$ denotes the vorticity magnitude); $Q$ or $\lambda_2$ has the dimension of $\omega^2$; $\lambda_{ci}$ and $R$ has the dimension of $\omega$. In order to make these criteria comparable and also to facilitate the discussion of equivalent thresholds, we regularize these criteria to obtain the corresponding nondimensional forms. Specifically, the criteria are first processed by certain power operations and then normalized by the corresponding root mean squares (RMSs) at a fixed wall-normal position. Normalizing the criteria by their RMSs is a popular processing technique used in order to relieve the influence of wall-normal variation of the vortex strength.[35] Note that only the data in the effective range of the identification criteria (see the second column of Table I) is processed. The data outside this range is directly set as zeros in the nondimensional forms. Explicit formulas for the regularization process are shown in the last column of Table I. As a convention, a symbol with a hat implies that it is the nondimensional form obtained by the regularization process.

TABLE I. A collection of definitions of various identification criteria regarding the vortex strength

| Criterion | Definition | Effective range | Relation with the complex eigenvalue of $\nabla \mathbf{u}$ $(\lambda_{cr} \pm i\lambda_{ci})$ | Nondimensional form (rms: root mean square for a given height) |
|---|---|---|---|---|
| $\Delta$ | The discrimination: $\Delta = (R/2)^2 + (Q/3)^3$, where $Q$ and $R$ are the second and the third invariants for $\nabla \mathbf{u}$ | $\Delta > 0$ | $\Delta = \dfrac{\lambda_{ci}^6}{27}[1 + 9(\dfrac{\lambda_{cr}}{\lambda_{ci}})^2]^2$ ($\lambda_{cr}$ is the real part of the complex eigenvalue of $\nabla \mathbf{u}$) | $\hat{\Delta} = \begin{cases} \Delta^{\frac{1}{6}}/\mathrm{rms}(\Delta^{\frac{1}{6}}) & \Delta > 0 \\ 0 & \Delta \leq 0 \end{cases}$ |
| $Q$ | The second invariant for $\nabla \mathbf{u}$, $Q = (\|\mathbf{\Omega}\|^2 - \|\mathbf{S}\|^2)/2$, where $\|\cdot\|$ denotes Frobenius norm; $\mathbf{\Omega} = (\nabla \mathbf{u} - \nabla \mathbf{u}^T)/2$, $\mathbf{S} = (\nabla \mathbf{u} + \nabla \mathbf{u}^T)/2$ | $Q > 0$ | $Q = \lambda_{ci}^2[1 - 3(\dfrac{\lambda_{cr}}{\lambda_{ci}})^2]$ | $\hat{Q} = \begin{cases} Q^{\frac{1}{2}}/\mathrm{rms}(Q^{\frac{1}{2}}) & Q > 0 \\ 0 & Q \leq 0 \end{cases}$ |
| $\lambda_2$ | The second largest eigenvalue for $\mathbf{\Omega}^2 + \mathbf{S}^2$, which is equal to $-\nabla\nabla p/\rho$ if the viscous and unsteady terms is neglected | $\lambda_2 < 0$ | $\lambda_2 = \lambda_{ci}^2\left[\left(\dfrac{\lambda_{cr}}{\lambda_{ci}}\right)^2 - 1\right]$ only if $\nabla \mathbf{u}$ is a normal tensor | $\widehat{\lambda_2} = \begin{cases} |\lambda_2|^{\frac{1}{2}}/\mathrm{rms}(|\lambda_2|^{\frac{1}{2}}) & \lambda_2 < 0 \\ 0 & \lambda_2 \geq 0 \end{cases}$ |
| $\lambda_{ci}$ | The imaginary part of the complex eigenvalue of $\nabla \mathbf{u}$ | $\lambda_{ci} > 0$ | $\lambda_{ci}$ | $\widehat{\lambda_{ci}} = \lambda_{ci}/\mathrm{rms}(\lambda_{ci})$ |
| $R$ | The twice of the minimum angular speed of the material lines in the plane perpendicular to $\mathbf{\Lambda}_r$ | $R > 0$ | $R = \omega_r - \sqrt{\omega_r - 4\lambda_{ci}^2}$ ($\omega_r$ is the vorticity component along $\mathbf{\Lambda}_r$) | $\hat{R} = R/\mathrm{rms}(R)$ |

Facilitated by the regularization process, the correlation degree of these criteria could be quantitatively evaluated by the overall correlation coefficients. For example, the overall correlation coefficient for $\hat{\Delta}$ and $\hat{Q}$ is defined as

$$c(\hat{\Delta}, \hat{Q}) = \frac{\int_\Omega \hat{\Delta}\hat{Q} d\Omega}{\sqrt{\int_\Omega \hat{\Delta}^2 d\Omega \int_\Omega \hat{Q}^2 d\Omega}}, \quad (1)$$

where $\Omega$ represents the spatial domain of the considered DNS segment. The statistically averaged correlation coefficients for any two criteria from Table I are displayed in Fig. 1 by a matrix of three-dimensional (3D) bars. It shows that these criteria are highly correlated, with the lowest coefficients of 0.79. One special criterion deserving more attention is $\lambda_{ci}$, which has coefficients of larger than 0.9 with all the other four criteria. The peculiarity of $\lambda_{ci}$ has also been noticed from Table I, where all the other criteria are linked to $\lambda_{ci}$ by explicit formulas. It seems that choosing $\lambda_{ci}$ is a conservative strategy if the superiority of any other criterion has not been established yet.

Choosing the threshold is always the most challenging part in the application of the identification criteria. A suitable threshold should lead to well-recognized vortex structures, without tangling too much to influence the isolation of individual structures. Jiménez[36] suggested that the threshold could be determined based on the 'percolation' transition of isolated structures,[37] which typically takes place for an isolated volume fraction of 5–10%. In this work, a volume fraction of 6% for recognized vortex structures is prescribed. The resulting thresholds for these criteria are collected in Table II. As we can see, these thresholds take values in the same order, which is owing to the preceding regularization process.

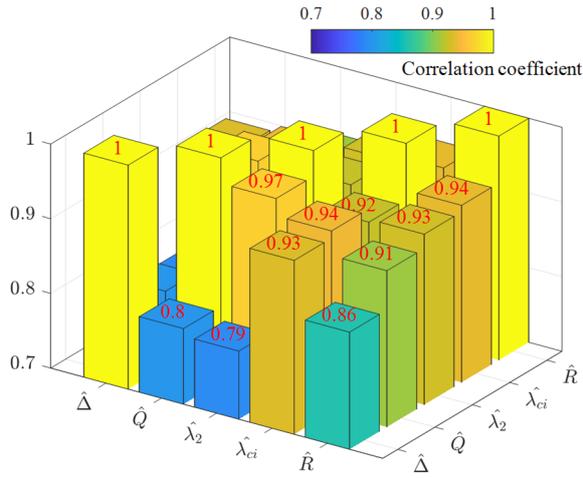

**FIG. 1**. A matrix of correlation coefficients for the five criteria

**TABLE II**. The thresholds of various criteria

| Criterion | $\hat{\Delta}$ | $\hat{Q}$ | $\widehat{\lambda_2}$ | $\widehat{\lambda_{ci}}$ | $\hat{R}$ |
|---|---|---|---|---|---|
| Threshold | 2.63 | 2.45 | 2.44 | 2.55 | 2.41 |

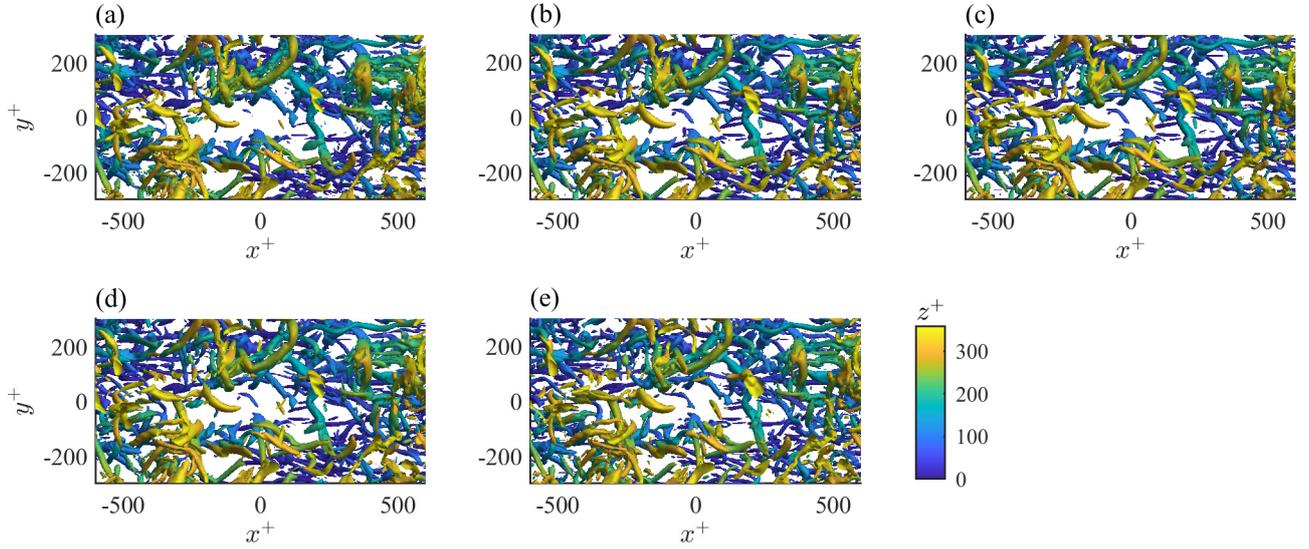

**FIG. 2**. The vortex structures identified by different criteria: (a) $\Delta$, (b) $Q$, (c) $\lambda_2$, (d) $\lambda_{ci}$, (e) $R$. The corresponding thresholds are listed in Table II. The isosurfaces are contoured based on the local wall-normal position ($z^+$) as shown by the colour bar at the lower right corner.

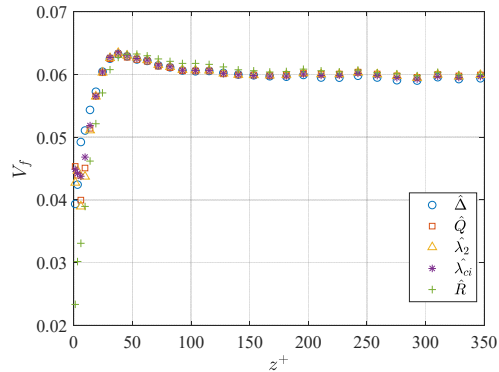

**FIG. 3**. Volume factions of identified vortices by different criteria as functions of wall-normal positions.

Fig. 2 shows the vortex structures identified by the five criteria using the thresholds listed in Table II. It shows that all the results present consistent tube-like vortex structures tangling with one another, which is a typical pattern for turbulent flows. The typical quasi-streamwise vortices and distorted hairpin-like vortices populate in the near-wall region and the logarithmic region (i.e. $z^+ > 80$), respectively. Furthermore, an increasing trend in the size of vortex tubes with the wall-normal position can also be observed, which is a characteristic of wall-bounded turbulent flows. Comparing the results from different criteria, we can recognize that the thickness and length of the same tubular vortex from different subplots are visually equivalent.

Quantitative comparison for the results of the five criteria is given in Fig. 3, where the volume factions for identified vortices are plotted against the wall-normal positions. It shows that with the increase of wall-normal positions, the volume faction first dramatically increases to the maximum peak, and then slowly reduce to a flat plateau of about 6%, which is equivalent to the prescribed value of the volume fraction. All the criteria give approximately equal volume factions for any given wall-normal positions, and the consistency is notably better for the logarithmic region. The results of Fig. 2 and Fig. 3 validate the equivalence of the thresholds for different criteria in the sense of isolating the same structures, which is an essential prerequisite for fairly comparing these results in the following discussions.

## C. Identification criteria for the vortex axis

As introduced in Sec. I, three adjoint vectors with the $\lambda_2$ or $\lambda_{ci}$ criteria are correlated with the vortex axis, which are the real eigenvector of $\nabla \mathbf{u}$ ($\mathbf{\Lambda}_r$), the normal vector of the plane spanned by the real part and imaginary part of the complex eigenvector ($\mathbf{\Lambda}_{cr} \times \mathbf{\Lambda}_{ci}$), and the eigenvector corresponding to the minimum eigenvalue of $\nabla \nabla p$ ($\mathbf{\Lambda}_p$). Unlike the criteria regarding the vortex strength, the relationships between these criteria for the vortex axis have not been sufficiently discussed in the published literatures. In this work, we will illustrate the physical meaning and the relationships of these criteria in an intuitive way. To begin, consider the relative velocity and pressure regarding one reference point, which has been labelled as vortex by both $\lambda_{ci}$ and $\lambda_2$. $\nabla \mathbf{u}$ contains all the information of the relative velocity. Specifically, for any relative position vector $\delta \mathbf{r}$ (infinitely small), the corresponding relative velocity is $\delta \mathbf{u} = \nabla \mathbf{u} \cdot \delta \mathbf{r}$ based on the definition of $\nabla \mathbf{u}$. The relative pressure $\delta p$ could be determined by a similar formula but with a higher-order truncation, which yields $\delta p = \nabla p \cdot \delta \mathbf{r} + \delta \mathbf{r} \cdot \nabla \nabla p \cdot \delta \mathbf{r}/2$. For the vortex core, particularly for the core centre, the pressure gradient in the vortex-axis-normal plane should be small, which indicates that it has a weak influence on the pattern of the local pressure distribution. Thus the term $\nabla p \cdot \delta \mathbf{r}$ can be neglected, which leads to $\delta p = \delta \mathbf{r} \cdot \nabla \nabla p \cdot \delta \mathbf{r}/2$. According to Jeong and Hussain[9], $\nabla \nabla p = -\rho(\mathbf{\Omega}^2 + \mathbf{S}^2)$ ($\rho$ is the density of the fluid) if the unsteady term and viscous terms are neglected. Thus, we can see both the relative velocity and pressure could be determined by $\nabla \mathbf{u}$.

Now take an example case of $\nabla \mathbf{u}$ as

$$\nabla \mathbf{u} = \begin{bmatrix} 1 & -3 & 0 \\ 2 & 1 & 0 \\ 1 & 1 & -2 \end{bmatrix}, \qquad (2)$$

The trace of $\nabla \mathbf{u}$ equals zero, indicating the situation of incompressible flow. The zero elements in the above matrix do not delimit the scope of the following discussion considering any 2-order tensor could be transformed into this form by only a coordinate rotation.[16] Fig. 4 shows the relative velocity and pressure on the surface of an infinitely small sphere centred at the reference point, with the sphere radius normalized as a unit. Note that only the tangential velocity components are displayed as vectors for clarity. The sphere surface shows two polar centres with zero tangential velocity which are located at the intersection points of the sphere surface and the third coordinate axis ($z$). The polar centres indicate the non-rotation axis ($\mathbf{\Lambda}_r$) according to Tian et al.[22]. Between the two polar centres, the velocity distribution represents a converging line and the vectors on the upper or lower sides tend to be tangential to it. This converging line forms a close circle on the sphere, which marks a plane of local swirling with zero off-plane velocities. Actually, the swirling plane is the plane spanned by the real part and the imaginary part of the complex eigenvector of $\nabla \mathbf{u}$. The corresponding normal direction (i.e. $\mathbf{\Lambda}_{cr} \times \mathbf{\Lambda}_{ci}$) has also been marked on the sphere surface. The distribution of relative pressure shows two minimum points on the sphere surface at the neighboring regions of the two polar centres. The minimum points correspond to the eigendirection of $\nabla\nabla p$ associated with the smallest eigenvalue, which is the definition of $\mathbf{\Lambda}_p$. Lastly, it should be pointed out that all of the three criteria correspond to two possible directions by adjusting the signs in their definitions. Following Gao et al.[13] and Tian et al.[16], only the direction with a smaller deviation angle with the local vorticity is used as the vortex axis direction. This convention is adopted in all the remaining parts of this article.

It seems that any one of the three criteria recognizes a privileged direction from a specific perspective, and all of them present some sense of rationality as the criteria of the vortex axis. Theoretically, the three recognized axes do not collapse unless $\nabla \mathbf{u}$ is a normal tensor, i.e. $\partial w / \partial x = \partial w / \partial y = 0$ in this example. Liu et al.[15] and Tian et al.[16] emphasized the physical interpretation of $\mathbf{\Lambda}_r$, and they defined the vortex vector by integrating the direction of $\mathbf{\Lambda}_r$ and the strength of $R$. However, in a loose sense, the vortex vector could be defined as any other combination of the foregoing criteria for the vortex strength and the vortex axis. This consideration necessitates revisiting the definition of the vortex vector by trialling the other possible combinations. A collection of all possible combinations of the criteria for the vortex strength and the vortex axis is displayed in Table III. A comprehensive comparison of these variants of vortex vectors will help to settle down the controversy on the most reasonable definition, which is meaningful for the foundation and application of the vortex identification criteria. The comparing works will be introduced in Sec. III and Sec. IV.

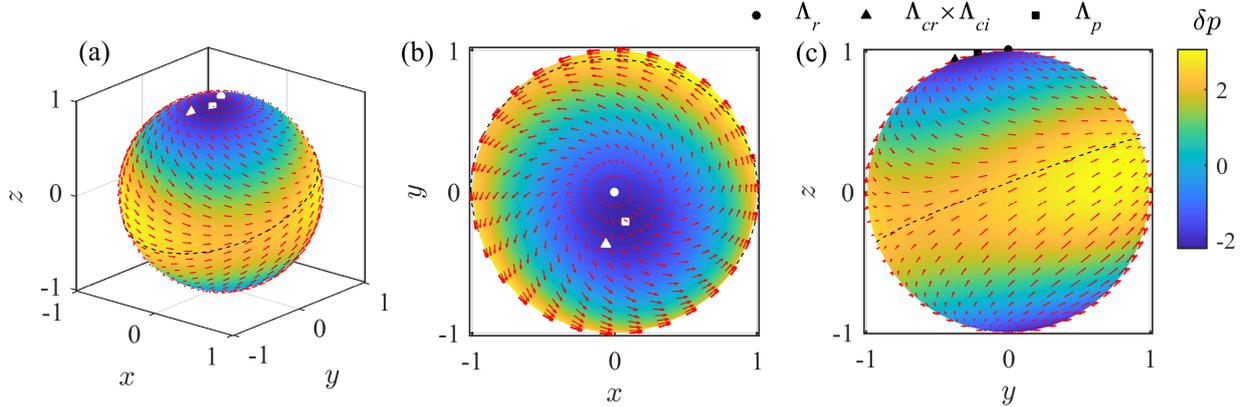

**FIG. 4**. The local relative velocity and pressure on an infinitely small sphere centred at a reference point, based on an example case of $\nabla \mathbf{u}$ as Eq. (2). The radius of the sphere has been normalized as a unit. Only the tangential velocity components are displayed as vectors for clarity. Three points marked by the small circle, triangle, and square represent the vortex axes recognized by $\mathbf{\Lambda}_r$, $\mathbf{\Lambda}_{cr} \times \mathbf{\Lambda}_{ci}$, and $\mathbf{\Lambda}_p$, respectively. The plane spanned by $\mathbf{\Lambda}_{cr}$ and $\mathbf{\Lambda}_{ci}$ is displayed by a large circle with the black dashed line.

TABLE III. A collection for various versions of the vortex vector

| Vortex axis \ Vortex strength | $\hat{\Delta}$ | $\hat{Q}$ | $\widehat{\lambda_2}$ | $\widehat{\lambda_{ci}}$ | $\hat{R}$ |
|---|---|---|---|---|---|
| $\mathbf{\Lambda}_r$ | V1-1 | V1-2 | V1-3 | V1-4 | V1-5 |
| $\mathbf{\Lambda}_{cr} \times \mathbf{\Lambda}_{ci}$ | V2-1 | V2-2 | V2-3 | V2-4 | V2-5 |
| $\mathbf{\Lambda}_p$ | V3-1 | V3-2 | V3-3 | V3-4 | V3-5 |

## III. THE EFFECTIVENESS REGARDING THE EFFICIENT COMPRESSION

Defining vortex as a vector form by integrating the criteria for the vortex strength and the vortex axis is necessary for completely describing the vortical motion and also for the decompression of the vortex representation. Generally, a 3D vector field contains three times as much as information as a 3D scalar field, which seems to violate the rule of the CR of wall-bounded turbulence. However, as demonstrated by Wang et al.[27], the vortex strength and the vortex axis are not wholly independent, and in fact, it is possible to infer the vortex axis by local curvatures of the isosurfaces. The novel aspect stems from the tube-like shape of vortex structures, which results in the good alignment of the vortex isosurface and the vortex axis.

We begin the analysis by showing the geometry of a typical vortex tube (see Fig. 5). Focus on a small surface element $S$ on the tube as marked by the red colour, and suppose the surface element is small enough to be approximated as a quadrant form. The normal unit vector of $S$ pointing to the outside is denoted as $\mathbf{n}$. According to the theory of differential geometry, two orthogonal directions can be determined on $S$, which correspond to the directions with the largest and the smallest curvatures, denoted as $\boldsymbol{\kappa}_1$ and $\boldsymbol{\kappa}_2$ respectively. For a typical tube-like geometry, particularly for a cylinder-shape one, it can be inferred that the two principal directions should be along the axis direction and along the azimuthal direction of the cross-section, as shown by $l_1$ and $l_2$ in the plot. Since $l_2$ bends towards the negative direction of $\mathbf{n}$, the corresponding curvature ($c_2$) is negative. The curve $l_2$ bends harder than the curve $l_1$, which means $|c_2| > |c_1|$, or $c_2 < c_1$ if the sign issue is considered. Thus, the vortex axis direction should be along the principal direction with the largest curvature ($c_1$), which is named as the first principle direction ($\boldsymbol{\kappa}_1$) in this work. This discussion reveals a possible route to infer the vortex axis direction based on local curvatures of the vortex isosurface.

The above analysis holds for ideal vortex tube models with comparatively straight vortex axis and uniform vortex radius. For the real situation, the bending of the vortex axis and the varying of the vortex radius may cause some deviations. From the perspective of vortex modelling, the alignment of $\boldsymbol{\kappa}_1$ and the vortex axis is undoubtedly beneficial since it will significantly reduce the number of variables required for describing a vortex field, indicating high compression efficiency. While $\boldsymbol{\kappa}_1$ is defined based on the isosurfaces of the identification criterion for the vortex strength (as listed in Table I), the vortex axis can be recognized by $\boldsymbol{\Lambda}_r$, $\boldsymbol{\Lambda}_{cr} \times \boldsymbol{\Lambda}_{ci}$, $\boldsymbol{\Lambda}_p$. We hope that the vortex structures or the vortex axis based on the above criteria should satisfy this novel property, which will be examined in the following analysis.

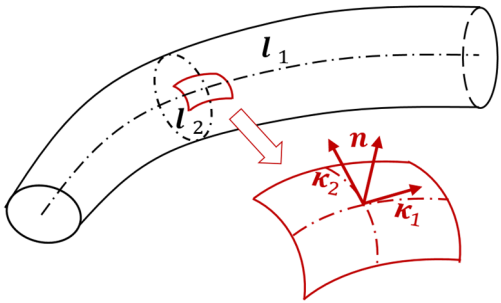

**FIG. 5**. A sketch for a vortex tube and the principal directions for local curvatures.

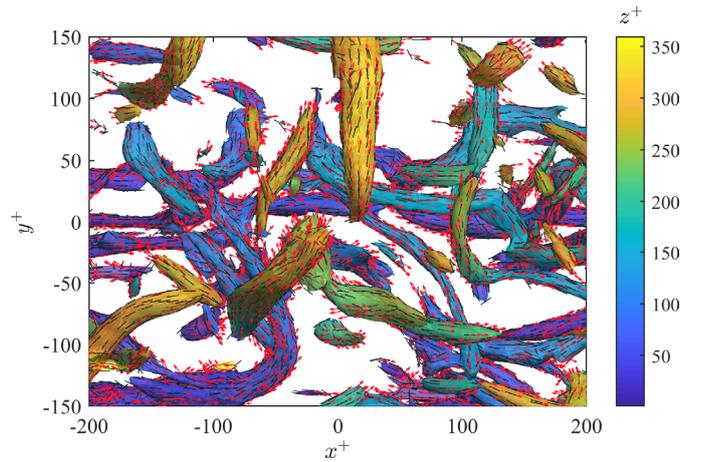

**FIG. 6**. An example vortex field showing the alignment of $\boldsymbol{\Lambda}_r$ (shown by red vectors) and the first principal direction of the local curvatures $\boldsymbol{\kappa}_1$ (displayed by black lines). The vortex isosurfaces identified by $\lambda_{ci}$ criterion are also displayed as references with contours showing the local wall-normal positions.

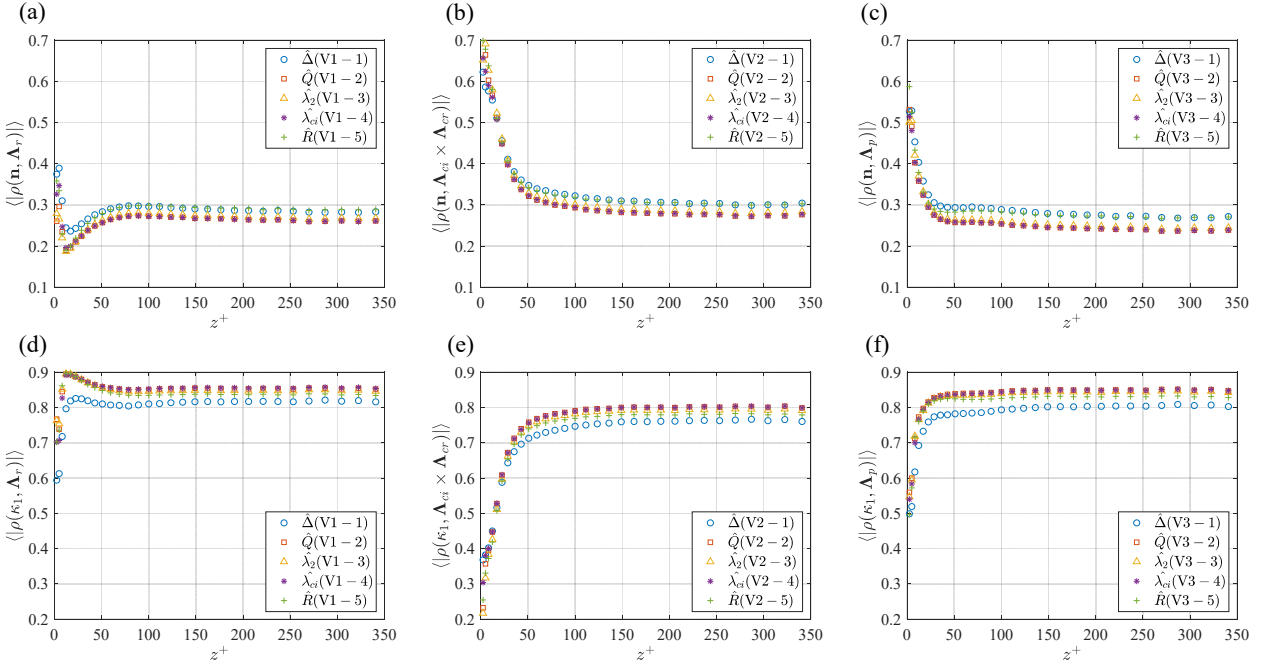

**FIG. 7**. Quantitative comparisons for the alignment of the vortex isosurface and the vortex axis based on various criteria (see Table III). The first row shows the correlation coefficients for the normal vector of the vortex isosurface ($\mathbf{n}$) and the recognized vortex axis (a-c: $\mathbf{\Lambda}_r$, $\mathbf{\Lambda}_{cr} \times \mathbf{\Lambda}_{ci}$, and $\mathbf{\Lambda}_p$). The second row shows the correlation coefficients for the first principal axes of local curvatures ($\mathbf{\kappa}_1$) and the recognized vortex axis (d-f: $\mathbf{\Lambda}_r$, $\mathbf{\Lambda}_{cr} \times \mathbf{\Lambda}_{ci}$, and $\mathbf{\Lambda}_p$).

As an example, Fig. 6 shows the isosurfaces of $\lambda_{ci}$ for the prescribed threshold with $\mathbf{\Lambda}_r$ and $\mathbf{\kappa}_1$ marked as vectors or line segments. Several interesting phenomena could be observed in the figure. Firstly, we can see that the vectors of $\mathbf{\Lambda}_r$ are tangential to the isosurface, namely $\mathbf{\Lambda}_r \cdot \mathbf{n} \approx 0$. Gao et al.[38] and Gao and Liu[20] attempted to explain the reason for this property based on the mathematical definition of $\mathbf{\Lambda}_r$. This property is also reminiscent of the vorticity-surface field (VSF) employed by Xiong and Yang[39] when visualizing vortex structures. VSF is defined as a scalar field with the gradient direction perpendicular to the local vorticity. By this definition, $\lambda_{ci}$ could be approximately regarded as the "vortex-surface field" if $\mathbf{\Lambda}_r$ is viewed as the local vortex direction. The more striking aspect is that the directions indicated by $\mathbf{\Lambda}_r$ and $\mathbf{\kappa}_1$ agree well for most cases, which is consistent with the foregoing analysis regarding the ideal vortex tubes. Particularly, longer and thinner vortex tubes represent better alignment for the two directions while the vortex blobs and the ends of vortex tubes see comparatively worse alignment. Considering it suffers from the irregularity of vortex tubes in turbulence, the alignment is overall satisfying.

To quantitatively describe the alignment of the vortex axis and the vortex isosurface, the pointwise correlation coefficients for $\mathbf{\Lambda}_r$ and $\mathbf{n}$ and for $\mathbf{\Lambda}_r$ and $\mathbf{\kappa}_1$ are calculated based on

$$\rho(\mathbf{\Lambda}_r, \mathbf{n}) = \frac{\mathbf{\Lambda}_r \cdot \mathbf{n}}{|\mathbf{\Lambda}_r||\mathbf{n}|}, \qquad (3)$$

$$\rho(\mathbf{\Lambda}_r, \mathbf{\kappa}_1) = \frac{\mathbf{\Lambda}_r \cdot \mathbf{\kappa}_1}{|\mathbf{\Lambda}_r||\mathbf{\kappa}_1|}. \qquad (4)$$

What is noteworthy is that $\rho(\cdot,\cdot)$ is different from the foregoing $c(\cdot,\cdot)$ in Eq. (1): the former means the pointwise correlation for two input signals while the later returns integral correlation coefficients in the considered domain. Considering that both $\mathbf{n}$ and $\mathbf{\kappa}_1$ might change directions by adding a negative sign in the definition, only the magnitude for $\rho(\mathbf{\Lambda}_r, \mathbf{n})$ or $\rho(\mathbf{\Lambda}_r, \mathbf{\kappa}_1)$ is considered, i.e. $|\rho(\mathbf{\Lambda}_r, \mathbf{n})|$ or $|\rho(\mathbf{\Lambda}_r, \mathbf{\kappa}_1)|$. While $|\rho(\mathbf{\Lambda}_r, \mathbf{n})|$ measures the adhesive degree of the vortex axis and the vortex isosurface, $|\rho(\mathbf{\Lambda}_r, \mathbf{\kappa}_1)|$ indicates whether the

recognized vortex axis is well aligned with the first principal direction of local curvatures. A good alignment of the vortex axis and the vortex isosurface promises small $|\rho(\mathbf{\Lambda}_r, \mathbf{n})|$ and large $|\rho(\mathbf{\Lambda}_r, \mathbf{\kappa}_1)|$.

Generally, for each point of the DNS data, there exits an isosurface of vortex crossing this point. Thus, the curvatures of local isosurfaces can be calculated on every grid points of the DNS data. The calculation for principal directions of local curvatures is not detailed in this article. The reader can also refer to the article of Wang et al.[27], who provided a detailed introduction about it in the appendix. All the grid points recognized as vortices by the prescribed threshold (see Table II) are counted in a statistical average of $|\rho(\mathbf{\Lambda}_r, \mathbf{n})|$ and $|\rho(\mathbf{\Lambda}_r, \mathbf{\kappa}_1)|$. The results are denoted as $\langle|\rho(\mathbf{\Lambda}_r, \mathbf{n})|\rangle$ and $\langle|\rho(\mathbf{\Lambda}_r, \mathbf{\kappa}_1)|\rangle$, with a bracket indicating the statistically averaging operation. Similarly, the calcualation and statistical process could be performed on other definitions of the vortex strength and the vortex axis, and the details are not repeated since the corresponding formulas are just similar.

Fig. 7 displays an ensemble of the average correlation coefficients for the vortex axis and $\mathbf{n}$ or $\mathbf{\kappa}_1$ based on the various definitions for the vortex vector (see Table III). It shows that the $\Delta$ criterion perform the worst in the alignment of the vortex isosurface and axis, with comparatively larger correlation coefficients for $\mathbf{n}$ and the recognized vortex axis, and smaller correlation coefficients for $\mathbf{\kappa}_1$ and the recognized vortex axis. The results of $Q$, $\lambda_2$, $\lambda_{ci}$ are very close for all the cases. The newly-proposed criterion $R$ does not perform very well in this examination, particularly for the first row of this figure where $R$ performs as bad as the $\Delta$ criterion. The unsatisfying performance of $R$ is not surprising since the advantage of this criterion is removing the influence of shear flow, which is not directly correlated to the alignment property. This numerical test demonstrates that $Q$, $\lambda_2$, $\lambda_{ci}$ are better criteria in defining a vortex field with good alignment, which is beneficial for the efficient compression of wall-bounded turbulence.

Besides comparing the five criteria regarding the vortex strength, a horizontal comparison shows that $\mathbf{\Lambda}_r$ criterion leads to the best alignment, whose superiority is more evident for the near-wall region. In this region, $\mathbf{\Lambda}_r$ criterion brings a maximum peak for the correlation coefficients for the vortex axis and $\mathbf{\kappa}_1$, while the results of $\mathbf{\Lambda}_{cr} \times \mathbf{\Lambda}_{ci}$ and $\mathbf{\Lambda}_p$ maintain at a low level. In fact, most of the vortices in the buffer layer are quasi-streamwise vortices, which take simple tube-like shapes compared to the irregular ones in the logarithmic region. The bad performance of $\mathbf{\Lambda}_{cr} \times \mathbf{\Lambda}_{ci}$ and $\mathbf{\Lambda}_p$ in this region might be caused by the intense shear layer embedded in this region. It implies that the two criteria are more susceptible to the influence of shear flow. Thus, it validates that the $\mathbf{\Lambda}_r$ criterion is better than the other two criteria for the vortex axis regarding the alignment of the vortex axis and the vortex isosurface.

## IV. THE EFFECTIVENESS REGARDING THE ACCURATE DECOMPOSITION

Decompressing the vortex representation of wall-bounded turbulence necessitates a V2V reconstruction. V2V reconstruction is an inverse problem for the vortex identification process. Although most researchers tend to employ Biot-Savart law to recover velocity fields based on the vortex filaments, the results are questionable since the law strictly holds for calculating the inducing velocity of vorticity rather than vortex. In fact, in the near-wall region, the orientation of the local vorticity and the vortex, and their organized patterns could be very different.[40, 41] Wang et al.[27] proposed a data-driven method to reconstruct the velocity field based on the vortex field, which was named as field-based stochastic estimation (FLSE). FLSE allows the estimation of one 3D field based on another field under the framework of the linear stochastic estimation. Rather than directly using the 3D FLSE developed by Wang et al.[27], this work confines this discussion in 2D V2V reconstruction for the $x - y$ plane. The reason is that a 3D V2V reconstruction requires much more calculation resources, which is not acceptable in this test since the number of testing cases is huge. Instead, 2D V2V reconstruction is simple in implementation and the corresponding results are also valuable to judge the effectiveness of these criteria. It is believed that the performances of these criteria in 3D V2V reconstruction can be inferred based on their performances in the 2D reconstruction for various the wall-normal positions.

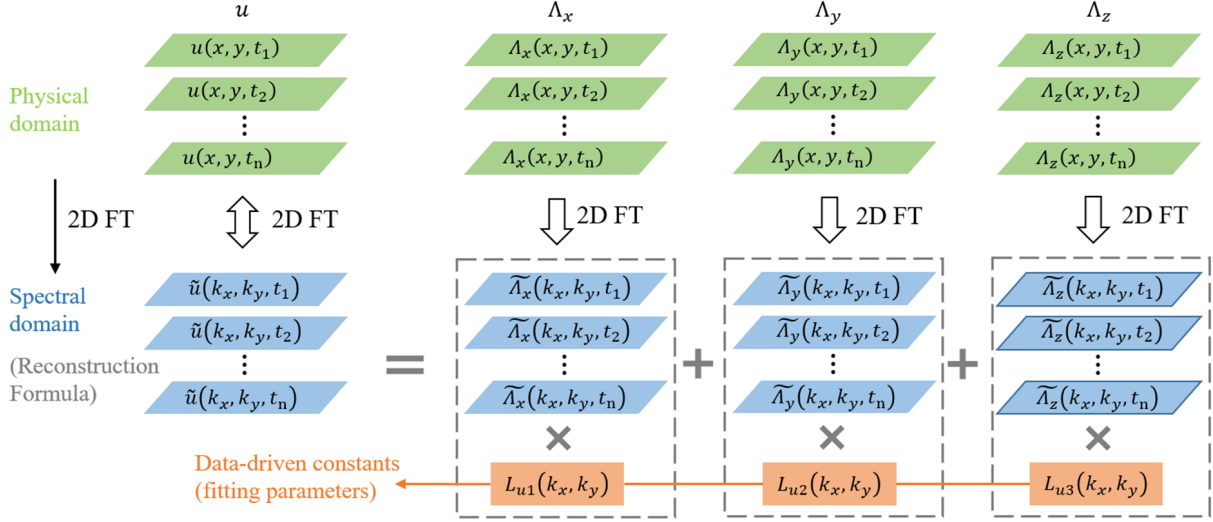

**FIG. 8**. A diagram of 2D V2V reconstruction of $u(x,y,t)$ via FLSE

Considering that wall-bounded turbulence shows the homogenous property in the $x-y$ plane, the 2D Fourier transform can be employed in the implementation of FLSE. In this situation, the 2D FLSE could be viewed as the 2D version of the spectral linear stochastic estimation (SLSE) developed by Tinney et al.[28]. To introduce the 2D FLSE, we take the reconstruction for $u(x,y,t)$ as an example. Note that the symbol $t$ labels different samples of the $u$ field, either obtained from different instantaneous times or truncated from other positions in the same homogenous plane. The 2D Fourier Transform (FT) of $u(x,y,t)$ in the $x-y$ plane is designated as $\tilde{u}(k_x,k_y,t)$, where $k_x, k_y$ denote the angular wavenumbers along the $x,y$ direction, respectively. For the convenience of the following discussion, let $\Lambda_x$, $\Lambda_y$ and $\Lambda_z$ denote the three components of the vortex vector, which might be obtained by any one of the definitions in Table III. The 2D FT of $\Lambda_x$, $\Lambda_y$ and $\Lambda_z$ are represented by $\widetilde{\Lambda_x}(k_x,k_y,t)$, $\widetilde{\Lambda_y}(k_x,k_y,t)$ and $\widetilde{\Lambda_z}(k_x,k_y,t)$, respectively.

Suppose that $\tilde{u}(k_x,k_y,t)$ could be estimated by the linear combination of $\widetilde{\Lambda_x}(k_x,k_y,t)$, $\widetilde{\Lambda_y}(k_x,k_y,t)$ and $\widetilde{\Lambda_z}(k_x,k_y,t)$, which yields

$$\tilde{u}(k_x,k_y,t) \sim \widetilde{\Lambda_x}(k_x,k_y,t)L_{u1}(k_x,k_y) + \widetilde{\Lambda_y}(k_x,k_y,t)L_{u2}(k_x,k_y) + \widetilde{\Lambda_z}(k_x,k_y,t)L_{u3}(k_x,k_y), \quad (5)$$

where $L_{u1}$, $L_{u2}$ and $L_{u3}$ are unknown constants independent of $t$. For fixed $k_x, k_y$, the equations with different $t$ are collected into a set of linear equations with the three constants unknown, which should be solved in a least-square manner. In this way, the constants could be viewed as the fitting parameters for the estimation formula of Eq. (5). When the constants $L_{u1}$, $L_{u2}$ and $L_{u3}$ are determined, Eq. (5) can be employed to estimate $\tilde{u}(k_x,k_y,t)$ based on $\widetilde{\Lambda_x}(k_x,k_y,t)$, $\widetilde{\Lambda_y}(k_x,k_y,t)$ and $\widetilde{\Lambda_z}(k_x,k_y,t)$. In this way, the velocity field can be estimated scale by scale in the Fourier space. After all the scales of $\tilde{u}(k_x,k_y,t)$ are estimated, an inverse FT is performed to return the reconstructed $u(x,y,t)$. For clarity, a diagram showing the road map of 2D FLSE regarding the reconstruction of $u(x,y,t)$ is provided in Fig. 8. The reconstruction for the $v$ and $w$ can be carried out by the same procedure.

To implement the V2V reconstruction, we consider a calculation domain of $\delta \times 0.5\delta$ in the $x-y$ plane. This aspect ratio gains support from the statistical results for attached structures.[35] Generally, the larger calculation domain will bring better reconstruction results, yet heavier calculation burden. In order to calculate the involved constants in Eq. (5), the DNS data segments were chopped into blocks of $\delta \times 0.5\delta$. Totally, 182 resulting blocks were employed to solve the estimation constants. Fig. 9 provides the contour map of the reconstructed velocity field based on the vortex field defined by V1-4 in Table III, compared with the original velocity field. It shows that the reconstructed velocity field is reasonably consistent with the original one. Particularly, the consistency for the $v$ or $w$ component is better than that for the $u$ component. It is because that $u$ component corresponds to large coherent scales, which might be beyond the considered calculation domain. In total, the reconstructed results are very promising since the input vortex field only contains the information of vortex filaments. Furthermore, it should be reminded that only the vortex field in the $x-y$ plane is considered in the reconstruction, although the real inducing effect of the vortex is three-dimensional. The success in this decompression procedure also validates that the vortex representation is an effective way to compress the wall-bounded turbulence.

To quantitatively describe the reconstruction accuracy, the overall correlation coefficient for the original velocity field and the reconstructed one is defined as

$$c(\mathbf{u}_{\text{DNS}}, \mathbf{u}_{\text{rec}}) = \sqrt{[c(u_{\text{DNS}}, u_{\text{rec}})^2 + c(v_{\text{DNS}}, v_{\text{rec}})^2 + c(w_{\text{DNS}}, w_{\text{rec}})^2]/3}, \qquad (6)$$

where $c(u_{\text{DNS}}, u_{\text{rec}})$, $c(v_{\text{DNS}}, v_{\text{rec}})$, $c(w_{\text{DNS}}, w_{\text{rec}})$ stand for the overall correlation coefficients for three velocity components field as defined by Eq. (1).

Fig. 10 shows the averaged correlation coefficient $\langle c(u_{\text{DNS}}, u_{\text{rec}}) \rangle$ for the original velocity fields and the reconstructed ones based on different definitions of vortex (as listed in Table III) as a function of the wall-normal position. For each vortex definition, two reconstruction cases are considered. In the first case, the vortex field defined by any item of Table III is directly reconstructed, which counts in the contributions of all the data points in the effective ranges of the criteria. In the second case, only intense vortices whose strength is larger than the threshold (see Table II) are considered in the reconstruction, which involves a threshold-filtering process. A comprehensive analysis on the results of the two reconstruction cases could reveal the sensitivity of the reconstruction accuracy to the threshold. The corresponding results for the two cases are shown in the first and second rows of Fig. 10, respectively, which is very revealing in several aspects. Firstly, in each subplot, it shows a clear decreasing trend of the reconstruction accuracy with the increase of $z^+$ above the buffer layer. As we know, the scales of the coherent structures increase with the wall-normal position, yet the width of the V2V reconstruction domain remains unchanged in this work. Therefore, larger parts of the coherent vortices are not accounted for in the reconstruction at the higher wall-normal position, which explains the dropping trend of the reconstruction accuracy.

Secondly, for the non-threshold case (the first row in Fig. 10), the $\Delta$ criterion leads to the highest reconstruction accuracy, closely followed by the $\lambda_{ci}$ criterion. The performances of the other three criteria lag far behind, compared to the former two criteria. The superiority of the $\Delta$ and $\lambda_{ci}$ criteria might be attributed as the comparatively looser requirement in recognizing vortices, which can be indicated by the formulas in the fourth column of Table I. In other words, the $Q$ and $\lambda_2$ criteria only recognized part of the vortices which have been recognized by the $\Delta$ or $\lambda_{ci}$ criterion. The $R$ criterion is an exception for this explanation since $R > 0$ are equivalent to $\Delta > 0$ or $\lambda_{ci} > 0$. By definition, the $R$ criterion is determined by the minimum angular speed of material lines in the vortex-axis-normal plane. By neglecting the influence of the maximum angular speed, the $R$ criterion has a good performance in removing the influence of the shear flow.[20] However, the local maximum angular speed is very important information for inferring the flow field, and missing the information causes larger biases in the reconstructed results.

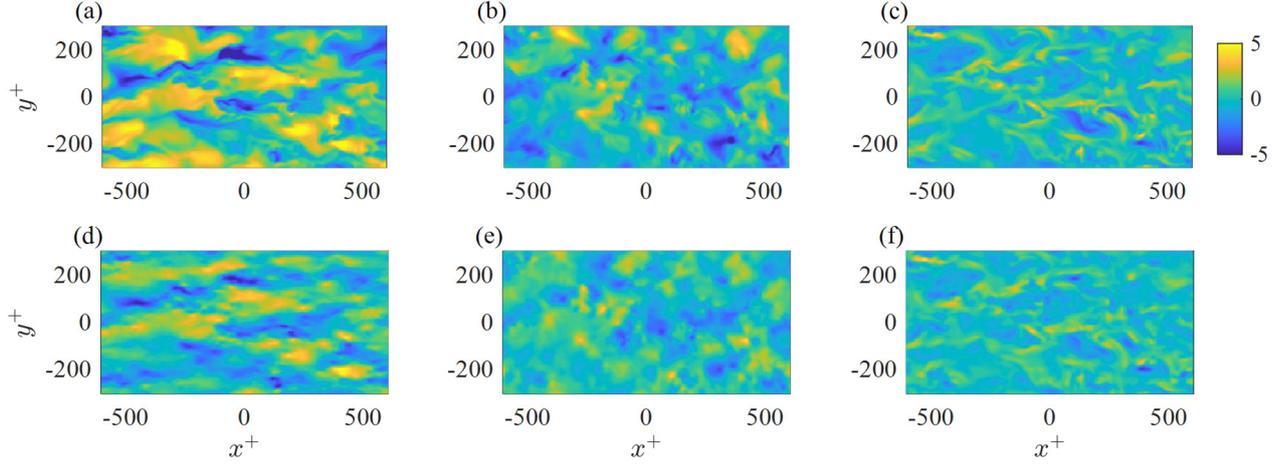

**FIG. 9**. Original DNS velocity fields (a-c) and the reconstructed velocity fields (d-f) based on the vortex field defined by V1-4. The section is extracted from $z^+ = 50.9$ and the three columns correspond to $u^+$, $v^+$ and $w^+$ respectively.

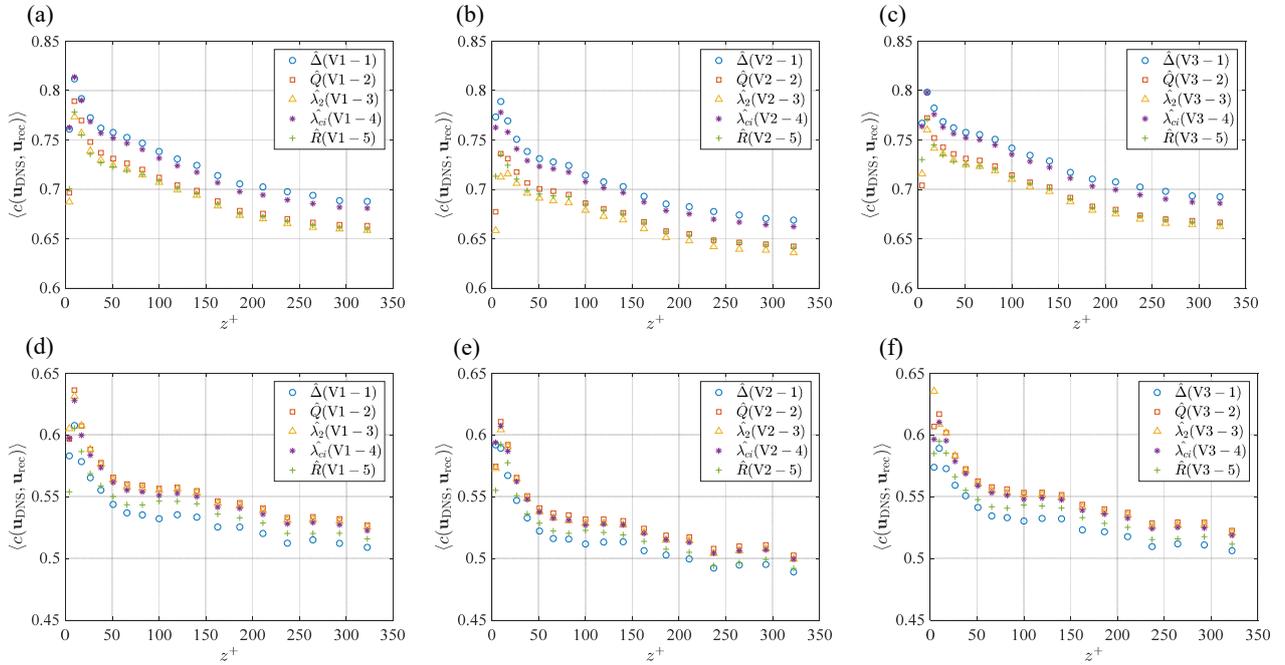

**FIG. 10**. Overall correlation coefficients for original velocity fields and the reconstructed ones based on different versions of vortex fields. Results of the first row correspond to the reconstruction based on the vortex vector fields defined by various criteria, without threshold filtering. Results of the second row correspond to the reconstruction based on intense vortex vectors obtained by a threshold filtering on the criteria. The three columns from left to right correspond the results based on the vortex axis of $\boldsymbol{\Lambda}_r$, $\boldsymbol{\Lambda}_{cr} \times \boldsymbol{\Lambda}_{ci}$, or $\boldsymbol{\Lambda}_p$.

When it comes to the case for the threshold-filtering reconstruction (the second row in Fig. 10), things are very different. The reconstruction accuracy is obviously lower for this case, which is expected since the threshold filtering keeps only 6% of the total volume with non-zero values. In this case, the $\Delta$ criterion performs the worst while $Q$ and $\lambda_2$ criteria achieve the best results. The performance of the $\lambda_{ci}$ is also satisfying because its results are very close to the best ones. Thus, it concludes that the $\lambda_{ci}$ criterion is a stable choice for the vortex identification with promising performances for both weak and strong vortices. Another advantage of the $\lambda_{ci}$ criterion is that it has the same dimension as vorticity, which means $\lambda_{ci}$ could be directly viewed as the vortex strength, without the regularization process as introduced in Sec. II.

On the other hand, a comparison among the results of various criteria for the vortex axis in Fig. 10 (along the row) indicates that the $\Lambda_r$ and $\Lambda_p$ criteria bring approximately equivalent reconstruction accuracy, while the $\Lambda_{cr} \times \Lambda_{ci}$ criterion causes comparatively poor accuracy. These results, together with the results in Sec. III validate that the $\Lambda_r$ criterion is the best candidate for the definition of the vortex axis among the three criteria considered, which supports the application of the $\Lambda_r$ criterion in wall-bounded turbulence.[13, 14]

## V. CONCLUSION

This work investigated the effectiveness of various vortex identification criteria regarding the CR of wall-bounded turbulence, which involves two essential aspects: the alignment of the vortex axis and the vortex isosurfaces and the accuracy for the decompressing process. Notably, five criteria for the vortex strength including $\Delta$, $Q$, $\lambda_2$, $\lambda_{ci}$, $R$ and three criteria for the vortex axis including $\Lambda_r$, $\Lambda_{cr} \times \Lambda_{ci}$ and $\Lambda_p$ were considered in this investigation.

We first discussed the relationship between these identification criteria. To facilitate the following comparison, we regularized the criteria for the vortex strength to get their nondimensional forms. The relationships and the correlation degrees for these criteria of vortex strength were analyzed. The thresholds were prescribed so that the isolated structures took a volume fraction of 6%. It showed that the recognized structures from various criteria resemble one another and the volume fractions as functions of wall-normal positions collapse well, which was essential for making a fair comparison. The criteria for the vortex axis were interpreted in an intuitive way, which indicated that each of the criteria recognized a privileged direction. The definition of vortex vector was revisited by combining various criteria for the vortex strength and the vortex axis.

The vortex tube with a simple geometry promises a good alignment of the vortex axis and the vortex isosurface, which facilitates the compressed representation of vortex fields. The alignment degree was quantitatively evaluated by using two quantities: the pointwise correlation coefficients for the vortex axis and the normal direction of the local vortex isosurface, and the pointwise correlation coefficients for the vortex axis and the first principal direction of local curvatures. It was shown that the vortex axis tended to be tangential to the local vortex isosurface and aligned with the first principal direction of local curvatures. The statistical results showed that while $Q$, $\lambda_2$, $\lambda_{ci}$ are better criteria for the vortex strength, $\Lambda_r$ performed the best in the three criteria for the vortex axis regarding the good alignment of the vortex axis and the vortex isosurface.

Evaluating the accuracy of the decomposed velocity fields necessitates a V2V reconstruction. The V2V reconstruction was implemented by 2D FLSE, which is a data-driven method based on the Fourier Transform. Promising reconstruction results were found by comparing the reconstructed velocity fields to the original DNS velocity fields. Quantitative comparison on the reconstruction accuracy based on various definitions of vortices validated that the $\lambda_{ci}$ criterion is a good choice for the vortex strength since its performance demonstrated promising results for both the non-threshold case and the threshold case. As for the criteria of the vortex axis, the $\Lambda_r$ and $\Lambda_p$ criteria performed equivalently well while the $\Lambda_{cr} \times \Lambda_{ci}$ criterion caused poor accuracy.

To collect all the results, we can conclude that a combination of the $\lambda_{ci}$ criterion for the vortex strength and the $\Lambda_r$ criterion for the vortex axis is a superior definition for vortex vector in the application of CR for wall-bounded turbulence. The evaluating principles employed in the current work also benefit the assessment works of vortex identification criteria in the application of other flows.

**ACKNOWLEDGEMENTS**

This work was supported by the National Natural Science Foundation of China (grant 11902371, 91852204) and the Fundamental Research Funds for the Central Universities (19lgpy303, 2019QNA4056).

**DATA AVAILABILITY STATEMENT**

The data that support the findings of this study are openly available at "https://torroja.dmt.upm.es/turbdata/blayers/high_re/", reference number [29, 30].